\documentclass[twocolumn]{IEEEtran}
\usepackage{color}
\usepackage{amsmath}
\usepackage{amssymb}
\usepackage{graphicx}
\IEEEoverridecommandlockouts{}
\usepackage{subfigure}
\usepackage{algorithm}
\usepackage{array}
\usepackage[noend]{algpseudocode}
\usepackage{caption}
\usepackage{amsthm}
\usepackage{cases}

\newtheorem{Theorem}{\textbf{Theorem}}

\newtheorem{Lemma}{\textbf{Lemma}}

\begin{document}

\title{Analytical Solution of Poisson's Equation \\ with Application to VLSI Global Placement}

\author{
\normalsize \mbox{Wenxing Zhu$^{1}$,}~\mbox{Zhipeng Huang$^{1}$,}~\mbox{Jianli Chen$^{1}$,}~~and~\mbox{Yao-Wen Chang$^{2,3}$}\\
       \normalsize $^{1}$Center for Discrete Mathematics and Theoretical Computer Science, Fuzhou University, Fuzhou 350108, China\\
    \normalsize $^{2}$Graduate Institute of Electronics Engineering, National Taiwan University, Taipei 106, Taiwan\\
       \normalsize $^{3}$Department of\ Electrical Engineering, National Taiwan University, Taipei 106, Taiwan\\
       \normalsize wxzhu@fzu.edu.cn; N155420004@fzu.edu.cn; jlchen@fzu.edu.cn;  ywchang@ntu.edu.tw
}

\maketitle

%


\begin{abstract}
Poisson's equation has been used in VLSI global placement for  describing the potential field caused by a given charge density distribution.
Unlike previous global placement methods that solve Poisson's equation numerically, in this paper, we provide an analytical solution of the equation to calculate the potential energy of an electrostatic system. The analytical solution is derived based on the separation of variables method and an exact density function to model the block distribution in the placement region, which is an infinite series and converges absolutely. Using the analytical solution, we give a fast computation scheme of Poisson's equation and develop an effective and efficient global placement algorithm called \emph{Pplace}. Experimental results show that our Pplace achieves smaller placement wirelength than ePlace and NTUplace3. With the pervasive applications of Poisson's equation in scientific fields, in particular, our effective, efficient, and robust computation scheme for its analytical solution can provide substantial impacts on these fields.
\end{abstract}


\section{Introduction}
\label{sec:introduction}

In VLSI physical design, the placement problem is to place circuit blocks into a fixed die such that no block overlaps with the others and some cost metric (e.g., wirelength) is optimized. As the technology node enters the deep nanometer era with billion-transistor integration, the performance of a placement engine becomes dominant on the overall quality of an EDA tool~\cite{RefEplace}. As a result, many placers have been developed recently. There are three major types of placement algorithms~\cite{RefEss}: annealing-based methods, partitioning-based methods, and analytical-based techniques. Recent studies show that analytical placers typically achieve the
best placement quality with good scalability.

In analytical placement, one key ingredient is the technique to reduce overlaps among blocks for obtaining an evenly distributed placement. Many overlap reduction methods have been proposed in literature for analytical placement, e.g., partitioning, cell shifting, frequency control, assignment, bell-shaped density control, Helmholtz density control, and Poisson density control. A detailed description of these methods can be found in~\cite{RefEss}. Among them, the Poisson density control was adopted in several leading placers, e.g., FDP~\cite{RefSFDP}, Kraftwerk~\cite{RefKW2}, mFAR~\cite{RefmFAR}, and ePlace~\cite{RefEplace}. 

ePlace~\cite{RefEplace} used the Poisson's equation during global placement and achieved the best wirelength among all academic placers for the ISPD 2005 \cite{Ref2005} and ISPD 2006 \cite{Ref2006} benchmarks.
It first treated each block as a positive charge and the block density as a potential energy constraint.
Then, ePlace used Poisson's equation to model the electric potential and field distribution by a given charge density distribution, where
the Neumann boundary condition and the compatibility condition were set according to placement characteristics.
Poisson's equation is a partial differential equation (PDE).
By solving this PDE using the spectral method with a fast Fourier transform (FFT),
ePlace can rapidly compute an electric potential and field distribution,
thus spreading blocks quickly while minimizing the wirelength.

Poisson's equation is a PDE commonly used in many fields such as electrostatics, computer science,
mechanical engineering, theoretical physics, astronomy, chemistry, and so on.
For example, Poisson's equation on the rectangular domain can model a finite element system with bicubic Hermite basis functions~\cite{RefFE}.
%
Poisson's equation has also been widely used in electronic design automation. In addition to placement, for example, the authors in~\cite{RefGPU} formulated a traditional linear system as a special two-dimensional Poisson equation for large-scale power grid networks. In nanoscale double-gate CMOS, by solving Poisson's and Schr$\ddot{o}$dinger's equations, the authors in~\cite{RefNDCMOS} obtained a fully physical model for leakage distribution.
By using 3D-Poisson's equation, the work~\cite{Ref3D1} presented an analytical subthreshold model for trigate MOSFETs with thin BOX.

Solution methods for Poisson's equation can be divided into two categories: analytical solutions and numerical solutions~\cite{RefPDE}.
An analytical solution of a PDE is exactly the PDE solution, which may not be a closed-form solution \cite{RefAna}. In contrast, a numerical solution is obtained by some numerical methods, e.g., finite element methods, numerical approximation, interpolation methods, and so on. Since a numerical solution only approximates the PDE solution, such a numerical solution inevitably incurs some numerical error, and needs convergence analysis. For a half-space eigenstrain problem, for example, many previous works were proposed to reduce  numerical error as much as possible; in contrast, an  analytical solution for a PDE was presented finally in~\cite{Refa1}. Generally, if an  analytical solution can be found for a PDE, then it is intrinsically superior to a numerical one on no numerical error, easy for fast computation and  application.

Existing placers based on the Poisson density control used numerical methods to solve the Poisson's equation. In Kraftwerk, the authors used the geometric multi-grid solver DiMEPACK~\cite{RefDiMEPACK} to solve the Poisson's equation. In ePlace, the authors computed an electric potential and field distribution based on a general fast Fourier transform (FFT) package~\cite{RefFFTL}. However, Poisson's equation with different boundary conditions has different characteristics, and thus its solutions are different. Generally, it is a great challenge to obtain an analytical solution of a PDE, which even could not be possible.

For VLSI placement, ePlace \cite{RefEplace} has shown great advantages in distributing blocks and optimizing solution quality simultaneously. A natural question arises: can we further reduce the numerical error in ePlace and improve its solution quality?  Based on the electrostatic system in ePlace, in this paper, we attempt to derive an analytical solution of Poisson's equation and provide an effective and efficient mathematical model for VLSI placement.  The major contributions of our work are summarized as follows:
\begin{itemize}

%
%
%

\item We directly solve Poisson's equation to obtain an analytical solution, which is an infinite series. We also prove the convergence of this infinite series. 

\item  Based on the analytical solution, we propose a fast computation scheme for Poisson's equation in  $O(n\log n)$ time, where $n$ is the number of circuit blocks. And then, we develop an effective and efficient global placer called \emph{Pplace} (\textbf{P}oisson's equation with \textbf{a}nalytical solution and also an extension of ePlace).

\item To validate the effectiveness, efficiency, and robustness of our Pplace placer,
      we performed comparative studies with two leading placers on wirelength
      optimization, ePlace and NTUplace3~\cite{RefNtu3}, based on the ISPD 2005~\cite{Ref2005} and ISPD 2006~\cite{Ref2006} benchmarks which
      were used to evaluate placers in the wirelength-driven contest.
      Experimental results show that our Pplace placer can achieve smaller placement wirelength than ePlace. By replacing the density control in NTUplace3, we further embed the fast computation of Poisson's equation into NTUplace3, and reduced its total wirelength by 11\%, a significant improvement never achieved before.

\item With the pervasive applications of Poisson's equation in scientific fields, in particular, our effective, efficient, and robust computation scheme for its analytical solution can provide substantial impacts on these fields.
\end{itemize}

It should be noted that, we evaluate our Pplace placer based on wirelength
and runtime because it is easier to see the effects of the core engine of
the examined algorithm using these fundamental placement metrics. Our
techniques are readily applicable to placement with other considerations such as
routability and timing.

The remainder of this paper is organized as follows. We present some preliminaries in Section~\ref{sec:problem_statement}. In Section~\ref{sec:analytical solution and its convergence}, we show our exact density control of the placement problem, derive an analytical solution of  Poisson's equation, and provide a fast computation scheme of the equation.
 Section~\ref{sec:our placement algorithm} describes our placement algorithm.
Experimental results are reported in Section~\ref{sec:experimental results}.
Finally, Section~\ref{sec:conclusion} gives some conclusions.


\section{Preliminaries} \label{sec:problem_statement}

In this section, we describe the placement problem and explain the application of Poisson's equation in ePlace.

\subsection{Problem Statement}

Given $n$ blocks and $r$ nets, the circuit of the VLSI placement problem can be modelled as a hypergraph $G(V, E)$, where blocks are denoted by vertices $V = {\{v_1, v_2, v_3, \ldots, v_n\}}$, and nets are denoted by hyperedges $E = {\{e_1, e_2, e_3, \ldots, e_r \}}$.  Let $(w_i, h_i)$ be the width and height of  block $v_i$, and $(x_i, y_i)$ be the coordinate of the center of the block, respectively, $i=1, 2, \cdots, n$. Let the placement region be $[0, W]\times [0, H]$. The VLSI placement problem intends to determine the best position for each block  such that no block overlaps with the others, and the total wirelength is minimized \cite{RefEss}:
 \begin{equation}\label{eq:placement}
\min  \ W(v)~~\textmd{s.t.} ~~\textmd{no overlaps among blocks},
\end{equation}
where $W(v)$ is the total wirelength, which is calculated by the half-perimeter wirelength (HPWL)~\cite{RefEss}.

\subsection{Poisson's Equation in ePlace}\label{review ePlace}

 ePlace models the placement problem as a two-dimensional electrostatic system.
 According to the positions of all blocks in the placement region, the distributions of electric potential $\psi(x,y)$ and the electric field $\boldsymbol{\xi}(x,y)$ are determined, where $\boldsymbol{\xi}(x, y)=(\xi_x,\xi_y)=-\nabla\psi(x,y)$.  Then, each block $i$ is seen as a positive charge, where the area $A_i$ of block $i$ is regarded as the electric quantity $q_i$ of  charge $i$. Let
$\psi _i=\psi(x_i, y_i)$ and $\boldsymbol{\xi} _i=\boldsymbol{\xi}(x_i, y_i)$ denote the electric potential and electric field at charge $i$, respectively. Then the electric force $\boldsymbol{F}_i=q_i\boldsymbol{\xi} _i$ leads the movement of charge $i$.
 Thus, the system potential energy is defined as $N(v)=\frac {1}{2}\sum_{i\in V}N_i$, where $N_i=q_i\psi _i$ represents the potential energy of charge $i$. Finally, ePlace changes the density control in the constraint of problem~(\ref{eq:placement}) to the system of  total potential energy $N(v)=0$.

Based on Gauss's law, a partial differential equation with Poisson's equation, boundary condition, and the compatibility condition in ePlace \cite{RefEplace} is used:
\begin{subnumcases}{}\label{equ:ePlace-PDE-a}
\bigtriangledown \bullet \bigtriangledown \psi (x,y)=-\rho (x,y), & \\ \label{equ:ePlace-PDE-b}
\hat{n}\bullet \bigtriangledown \psi (x,y)=0,(x,y)\in \partial \mathbf{R},&\\ \label{equ:ePlace-PDE-c}
\iint_\mathbf{R}\rho (x,y)dxdy=\iint_\mathbf{R}\psi (x,y)dxdy=0.&
\end{subnumcases}
Equation (\ref{equ:ePlace-PDE-a}) gives Poisson's equation, where $\rho(x, y)$ is the density function.
Equation (\ref{equ:ePlace-PDE-b}) is the Neumann boundary condition, where
$\partial \mathbf{R}$ and $\hat{n}$ denote the boundary and the outer normal vector of the design range $\mathbf{R}$, respectively.
Equation (\ref{equ:ePlace-PDE-b}) is used for preventing a block from running out of the boundary.
Equation (\ref{equ:ePlace-PDE-c}) is the compatibility condition, which makes the system of equations a unique solution.

By dividing the placement region into $m \times m$ bins, ePlace uses local smoothness to get an $m\times m$ density matrix $\rho$, in which each element is calculated using the total area of blocks covered by the corresponding bin \cite{RefEplace}. Let $u$ and $p$ denote integer indexes, and the frequency components are defined as $\omega_{u}=\frac{2\pi u}{m}$ and $\omega_{p}=\frac{2\pi p}{m}$, respectively. The coefficient of each wave function of a discrete cosine transform is denoted by $a_{u,p}$ as follows:
  \begin{equation}
    a_{u,p}=\frac{1}{m^2}\sum _{l=0}^{m-1}\sum _{j=0}^{m-1}\rho(l,j)\cos(\omega_{u}l)\cos(\omega_{p}j).
    \label{equ:ePlace-a}
\end{equation}
Then, the potential value is calculated by
  \begin{equation}
    \psi(l,j)=\sum _{u=0}^{m-1}\sum _{p=0}^{m-1}\frac{a_{u,p}}{\omega_{u}^2+\omega_{p}^2}\cos(\omega_{u}l)\cos(\omega_{p}j),
    \label{equ:ePlace-psi}
\end{equation}
where $\psi(l,j)$ is the electric potential at the center of bin $b_{lj}$. In ePlace \cite{RefEplace},  $\boldsymbol{\xi}(l,j)$ is calculated as follows:
\begin{equation}
\boldsymbol{\xi}(l,j)=
  \left\{
   \begin{aligned}
   &\xi _{x}=\sum _{u=0}^{m-1}\sum _{p=0}^{m-1}\frac{a_{u,p}\omega_{u}}{\omega_{u}^2+\omega_{p}^2}\sin(\omega_{u}l)\cos(\omega_{p}j),  \\
   &\xi _{y}=\sum _{u=0}^{m-1}\sum _{p=0}^{m-1}\frac{a_{u,p}\omega_{p}}{\omega_{u}^2+\omega_{p}^2}\cos(\omega_{u}l)\sin(\omega_{p}j).
   \end{aligned}
   \label{equ:ePlace-ksi}
   \right.
  \end{equation}
Equations  \eqref{equ:ePlace-a}, \eqref{equ:ePlace-psi} and \eqref{equ:ePlace-ksi} are used in the global placement stage of ePlace. After global placement, detail placement and legalization are called to obtain a final placement result.

\section{Analytical Solution of Poisson's equation} \label{sec:analytical solution and its convergence}

 In this section, we first derive an analytical solution of the Poisson's equation. Then, we prove the convergence of the analytical solution and analyze its time complexity. Based on the analytical solution, finally, we propose a fast computation scheme of Poisson's equation.
\subsection{Analytical Solution of Poisson's Equation}\label{3b}

Similar to ePlace~\cite{RefEplace}, we have the continuous Poisson's equation, the relative boundary condition and the compatibility condition
\begin{subnumcases}{}\label{equ:PDE-a}
   \bigtriangledown \bullet \bigtriangledown \psi (\hat{x},\hat{y})=-\rho (\hat{x},\hat{y}),&  \\\label{equ:PDE-b}
   \hat{n}\bullet \bigtriangledown \psi (\hat{x},\hat{y})=0,(\hat{x},\hat{y})\in \partial \mathbf{R},&\\\label{equ:PDE-c}
   \iint_\mathbf{R}\psi (\hat{x},\hat{y})d\hat{x}d\hat{y}=0.&
\end{subnumcases}
In this subsection, we give an analytical solution of Poisson's equation with the relative boundary condition and the compatibility condition (\ref{equ:PDE-a})--(\ref{equ:PDE-c}).

By separation of variables, the density function in Equation (\ref{equ:PDE-a}) can be expanded as
\begin{equation}
\rho (\hat{x},\hat{y})=\sum _{u=0}^{\infty}\sum _{p=0}^{\infty}b_{u,p}X_u(\hat{x})Y_p(\hat{y}),
\label{td:density}
\end{equation}
and the potential can be calculated by
\begin{equation}
\psi (\hat{x},\hat{y})=\sum _{u=0}^{\infty}\sum _{p=0}^{\infty}a_{u,p}X_u(\hat{x})Y_p(\hat{y}).
\label{td:psi}
\end{equation}

According to Equations (\ref{equ:PDE-a}), (\ref{equ:PDE-b}), (\ref{td:density}) and (\ref{td:psi}), we have
\begin{equation}
  \left\{
   \begin{aligned}
   &\frac{\partial^{2}X_u(\hat{x})Y_p(\hat{y})}{\partial \hat{x}^{2}}+\frac{\partial^{2}X_u(\hat{x})Y_p(\hat{y})}{\partial \hat{y}^{2}}=-\frac{b_{u,p}}{a_{u,p}}X_u(\hat{x})Y_p(\hat{y}),  \\
   &\frac{\partial a_{u,p}X_u(\hat{x})Y_p(\hat{y})}{\partial \hat{x}}|_{\hat{x}=0,W}=0,\\
   &\frac{\partial a_{u,p}X_u(\hat{x})Y_p(\hat{y})}{\partial \hat{y}}|_{\hat{y}=0,H}=0.
   \end{aligned}
   \right.
   \label{td:PDE}
  \end{equation}
If Equations (\ref{td:PDE}) are satisfied, $\psi (\hat{x},\hat{y})$ is the solution of Poisson's equation.
We set $a_{u,p}\neq 0$ and $\alpha_{u,p}=-\frac{b_{u,p}}{a_{u,p}}$, thus
\begin{subnumcases}{}\label{td:1-a}
  X_u^{''}(\hat{x})Y_p(\hat{y})+X_u(\hat{x})Y_p^{''}(\hat{y})=\alpha_{u,p}X_u(\hat{x})Y_p(\hat{y})\hspace{1cm} &  \\\label{td:1-b}
   \frac{\partial X_u(\hat{x})Y_p(\hat{y})}{\partial \hat{x}}|_{\hat{x}=0,W}=\frac{\partial X_u(\hat{x})Y_p(\hat{y})}{\partial \hat{y}}|_{\hat{y}=0,H}=0&
\end{subnumcases}

We set $\frac{X_u^{''}}{X_u}=-(\frac{Y_p^{''}}{Y_p}+\alpha_{u,p})=-\beta_{u,p}^{2}, \beta_{u,p} >0$, thus we have
\begin{subnumcases}{}\label{td:2-a}
   X_u^{''}+\beta_{u,p}^{2}X_u=0,&  \\\label{td:2-b}
   X_u^{'}(0)=X_u^{'}(W)=0,~~0\leq\hat{x}\leq W.&
\end{subnumcases}
and
\begin{subnumcases}{}\label{td:3-a}
   Y_p^{''}+(\alpha_{u,p}-\beta_{u,p}^{2})Y_p=0,&  \\\label{td:3-b}
   Y_p^{'}(0)=Y_p^{'}(H)=0,~~0\leq\hat{y}\leq H.&
\end{subnumcases}

The solution of Equation (\ref{td:2-a}) is
\begin{subnumcases}{}\label{td:2-a-1}
   X_{u}(\hat{x})=C_{1}cos(\beta_{u,p}\hat{x})+C_{2}sin(\beta_{u,p}\hat{x}),&  \\\label{td:2-a-2}
   X_{u}^{'}(\hat{x})=-C_{1}\beta_{u,p}sin(\beta_{u,p}\hat{x})+C_{2}\beta_{u,p}cos(\beta_{u,p}\hat{x}).\hspace{0.7cm}&
\end{subnumcases}
According to the boundary conditions (\ref{td:2-b}), we can get $\beta_{u,p}=\frac{u\pi}{W}$ and $X_{u}(\hat{x})=C_{1}cos(\frac{u\pi}{W}\hat{x})$.


Similarly, the solution of Equation (\ref{td:3-a}) is $Y_{p}(\hat{y})=D_{1}cos(\frac{p\pi}{H}\hat{y})$.
According to the superposition principle, the solution of the original problem is
\begin{equation}
\psi (\hat{x},\hat{y})=\sum _{u=0}^{\infty}\sum _{p=0}^{\infty}a_{u,p}\cos(\frac {u\pi}{W}\hat{x})\cos(\frac {p\pi}{H}\hat{y}).
\label{equ:psi}
\end{equation}
Unlike Equation~(\ref{equ:ePlace-psi}), Equation~(\ref{equ:psi}) is a global function which can be used to calculate exactly the potential value of any point.

In order to calculate the coefficients $a_{u,p}$, we substitute Equation (\ref{equ:psi}) into Poisson's Equation (\ref{equ:PDE-a}).
By calculating $\bigtriangledown \bullet \bigtriangledown \psi (\hat{x},\hat{y})$, we can get another expression of the density function $\rho(\hat{x},\hat{y})$ as

\vspace{-0.2cm}
\begin{equation}\label{eq:p_xy}
    \small
        \begin{aligned}
    \rho(\hat{x},\hat{y}) = & \sum\limits^{\infty}_{p=1}\frac{p^2\pi^2}{H^2}a_{0,p}\cos(\frac{p\pi \hat{y}}{H})+\sum\limits^{\infty}_{u=1}\frac{u^2\pi^2}{W^2}a_{u,0}\cos(\frac{u\pi \hat{x}}{W})\\
    &+\sum\limits^{\infty}_{u=1}\sum\limits^{\infty}_{p=1}(\frac{u^2\pi^2}{W^2}+\frac{p^2\pi^2}{H^2})a_{u,p}\cos(\frac{u\pi \hat{x}}{W})\cos(\frac{p\pi \hat{y}}{H}).\\
    \end{aligned}
\end{equation}
Multiplying $\cos(\frac{\mu\pi}{W}\hat{x})\cos(\frac{\eta\pi}{H}\hat{y})$ ($\mu,\eta\geq0$) to both sides of Equation (\ref{eq:p_xy}), and making integration, we have
\begin{equation}
    \begin{aligned}
    &~~~~\iint_R\rho(\hat{x},\hat{y})\cos(\frac{\mu\pi}{W}\hat{x})\cos(\frac{\eta\pi}{H}\hat{y})d\hat{x}d\hat{y}\\
    &=\sum_{p=1}^\infty \frac{p^2\pi^2}{H^2}a_{0,p}\iint_R\cos(\frac{\mu\pi}{W}\hat{x})\cos(\frac{p\pi}{H}\hat{y})\cos(\frac{\eta\pi}{H}\hat{y})d\hat{x}d\hat{y}\\
    &+\sum_{u=1}^\infty \frac{u^2\pi^2}{W^2}a_{u,0}\iint_R\cos(\frac{u\pi}{W}\hat{x})\cos(\frac{\mu\pi}{W}\hat{x})\cos(\frac{\eta\pi}{H}\hat{y})d\hat{x}d\hat{y}\\
    &+\sum_{u=1}^\infty \sum_{p=1}^\infty (\frac{u^2\pi^2}{W^2}+\frac{p^2\pi^2}{H^2})a_{u,p}\iint_R\cos(\frac{u\pi}{W}\hat{x})\cos(\frac{\mu\pi}{W}\hat{x})\\
    &\cos(\frac{p\pi}{H}\hat{y})\cos(\frac{\eta\pi}{H}\hat{y})d\hat{x}d\hat{y}.
    \end{aligned}
    \label{equ:coscos}
\end{equation}

In Equation (\ref{equ:coscos}), the integration area $R=(0, W)\times (0, H)$. Hence, in the right-hand-side of Equation (\ref{equ:coscos}), by the orthogonality of the trigonometric functions, the first term takes a none-zero value only at $\mu=0$ and $p=\eta$, the second term only at $\eta=0$ and $u=\mu$,  and the third term only at $p=\eta$ and $u=\mu$.

Thus for $\mu\geq1$ and $\eta=0$, Equation (\ref{equ:coscos}) reduces to
\begin{equation*}
 \iint_R\rho(\hat{x},\hat{y})\cos\frac{\mu\pi \hat{x}}{W}d\hat{x}d\hat{y}=\frac{\mu^2\pi^2}{W^2}a_{\mu,0} \iint_R\cos^2\frac{\mu\pi \hat{x}}{W}d\hat{x}d\hat{y},
\end{equation*}
and we get
\begin{equation*}
a_{\mu,0}=\frac{2W}{\mu^2\pi^2H}\iint_R\rho(\hat{x},\hat{y})\cos(\frac{\mu\pi}{W}\hat{x})d\hat{x}d\hat{y}.
\end{equation*}

Similarly, we can get the coefficients of $a_{0,\eta}$ and $a_{\mu,\eta}$. To satisfy Equation (\ref{equ:PDE-c}), by Equation \eqref{equ:psi}, we have $a_{0,0} = 0$. Thus these coefficients can be summarized as follows:
\begin{subnumcases}{}\label{equation:a-a}
a_{0,0}=0,~~~~~~~~~~~~~~~~~~~~~~~~~~~~~~~~~~~~~~~~~~~~~~~~~  \\\label{equation:a-b}
a_{0,\eta}=\frac{2H}{\eta^{2}\pi^{2}W}\iint_\mathbf{R}\rho(\hat{x},\hat{y})\cos\frac{\eta\pi \hat{y}}{H}d\hat{x}d\hat{y},~(\eta\geq 1)\hspace{0.7cm}\\\label{equation:a-c}
a_{\mu,0}=\frac{2W}{\mu^{2}\pi^{2}H}\iint_\mathbf{R}\rho(\hat{x},\hat{y})\cos\frac{\mu\pi \hat{x}}{W}d\hat{x}d\hat{y},~(\mu\geq 1)\\
a_{\mu,\eta}=\frac{4WH}{(\mu^{2}H^{2}+\eta^{2}W^{2})\pi^{2}}\cdot\notag~~~~~~~~~(\mu\geq 1, \eta\geq 1)\\\label{equation:a-d}
~~~~~~~\iint_\mathbf{R}\rho(\hat{x},\hat{y})\cos\frac{\mu\pi \hat{x}}{W}\cos\frac{\eta\pi \hat{y}}{H}d\hat{x}d\hat{y}.
\end{subnumcases}

In order to make the Equations (\ref{equation:a-a})--(\ref{equation:a-d}) computable, we need a separable global density function.
Therefore, we introduce an exact density function $\rho (\hat{x},\hat{y})$ for the VLSI placement problem.
 Define the density function of a block $i$ in the $x$-direction as
\begin{equation*}
\theta _i(\hat{x})=
  \left\{
   \begin{aligned}
   &0,~~~~~&&|\hat{x}-x_i|>\frac {w_i}{2},  \\
   &1,~~~~~&&|\hat{x}-x_i|\leq \frac {w_i}{2}.
   \end{aligned}
   \right.
  \end{equation*}
Similarly, we can define the density function of block $i$ in the $y$-direction as $\theta _i(\hat{y})$. Then the density function of block $i$  is $\theta_i(\hat{x})\theta_i(\hat{y})$, and the density function of all blocks is
\begin{equation}
\rho (\hat{x},\hat{y})=\sum _{i=1}^{n}\rho_i (\hat{x},\hat{y})=\sum _{i=1}^{n}\theta _i(\hat{x})\theta _i(\hat{y}),
\label{equ:global density}
\end{equation}
where $n$ is the number of blocks.

 In order to satisfy Equation (\ref{equ:PDE-c}), we redefine $\rho (\hat{x},\hat{y})$ as
\begin{equation}
\begin{aligned}
\rho (\hat{x},\hat{y})&\triangleq\rho (\hat{x},\hat{y})-\frac{1}{WH}\iint_\mathbf{R}\rho (\hat{x},\hat{y})d\hat{x}d\hat{y}\\
&=\sum _{i=1}^{n}\theta _i(\hat{x})\theta _i(\hat{y})-\frac{1}{WH}\sum _{i=1}^{n}w_{i}h_{i}.
\end{aligned}
\label{equ:rho}
\end{equation}

Furthermore, for $\mu,\eta\geq 1$, we substitute Equation (\ref{equ:rho}) into Equation (\ref{equation:a-d}), and get
\begin{equation}
\begin{aligned}
&a_{\mu,\eta}\\
=&\frac{4WH}{(\mu^{2}H^{2}+\eta^{2}W^{2})\pi^{2}}(\sum _{i=1}^{n}\iint_R\rho_i (\hat{x},\hat{y})\cos\frac{\mu\pi \hat{x}}{W}\cos\frac{\eta\pi \hat{y}}{H}d\hat{x}d\hat{y} \\
&~~~-\frac{1}{WH}\sum _{i=1}^{n}w_{i}h_{i}\int^{W}_{0}\cos\frac{\mu\pi \hat{x}}{W}d\hat{x}\int^{H}_{0}\cos\frac{\eta\pi \hat{y}}{H}d\hat{y})\\
=&\frac{4W^2H^2}{(\mu^{2}H^{2}+\eta^{2}W^{2})\mu\eta\pi^{4}}\sum_{i=1}^{n}(\sin\frac {\mu\pi(x_i+\frac {w_i}{2})}{W}\\
&-\sin\frac {\mu\pi(x_i-\frac {w_i}{2})}{W})\cdot (\sin\frac {\eta\pi(y_i+\frac {h_i}{2})}{H}-\sin\frac {\eta\pi(y_i-\frac {h_i}{2})}{H}).
\end{aligned}\label{equ:aup}
\end{equation}

Similarly, we have
\begin{equation}
a_{\mu,0}=\frac{2W^2}{\mu^3\pi^3H}\sum_{i=1}^n h_{i}\left(\sin\frac{\mu\pi(x_{i}+\frac{w_{i}}{2})}{W}-\sin\frac{\mu\pi(x_{i}-\frac{w_{i}}{2})}{W}\right),
\label{equ:au0}
\end{equation}
and
\begin{equation}
a_{0,\eta}=\frac{2H^2}{\eta^3\pi^3W}\sum_{i=1}^n w_{i}\left(\sin\frac{\eta\pi(y_{i}+\frac{h_{i}}{2})}{H}-\sin\frac{\eta\pi(y_{i}-\frac{h_{i}}{2})}{H}\right).
\label{equ:a0p}
\end{equation}

Note that $a_{u,p}$ is calculated by the integral of the exact density function (\ref{equ:rho}) of the VLSI placement problem, which is more accurate than the discrete calculation in ePlace.

By Gauss's law, the electric field $\boldsymbol{\xi}(\hat{x},\hat{y})$ equals the negative gradient of the potential function $\psi (\hat{x},\hat{y})$.
 Based on  $\psi (\hat{x},\hat{y})$ in Equation~(\ref{equ:psi}), we have $\boldsymbol{\xi} (\hat{x},\hat{y})=(\xi _{\hat{x}},\xi _{\hat{y}})$, where
\begin{equation}
  \left\{
   \begin{aligned}
   &\xi _{\hat{x}}=\sum _{u=0}^{\infty}\sum _{p=0}^{\infty}\frac {u\pi}{W}a_{u,p}\sin(\frac {u\pi}{W}\hat{x})\cos(\frac {p\pi}{H}\hat{y}),  \\
   &\xi _{\hat{y}}=\sum _{u=0}^{\infty}\sum _{p=0}^{\infty}\frac {p\pi}{H}a_{u,p}\cos(\frac {u\pi}{W}\hat{x})\sin(\frac {p\pi}{H}\hat{y}).
   \end{aligned}
   \right.
   \label{equ:ksi}
  \end{equation}

From the methodology viewpoint, the major difference between our continuous analytical solution and the discrete solution in ePlace is that, ePlace uses the spectral method and the general fast Fourier transform to solve the PDE in Equations (\ref{equ:ePlace-PDE-a})--(\ref{equ:ePlace-PDE-c}), while we directly obtain an analytical solution of the PDE in Equations (\ref{equ:PDE-a})--(\ref{equ:PDE-c}).

\begin{figure}[!htb]
\centering
\subfigure[Original blocks]{
\label{FigA1a}
\includegraphics[width=40mm]{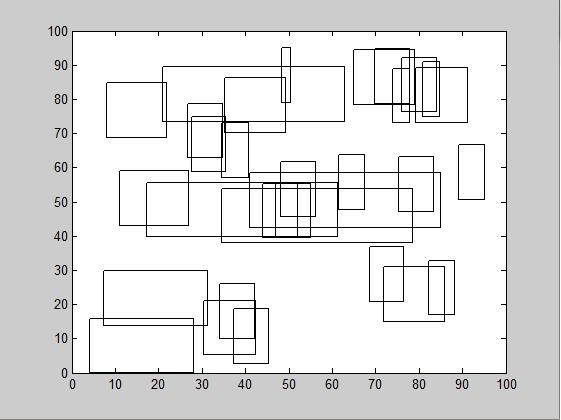}}
\subfigure[Original density distribution]{
\label{FigA1b}
\includegraphics[width=40mm]{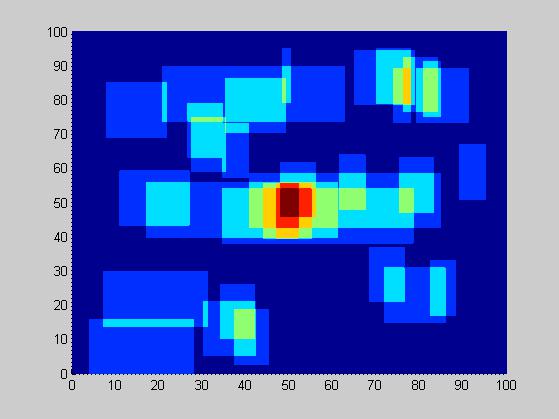}}
\subfigure[Analytical solution]{
\label{FigA1c}
\hspace{0.0cm}
\includegraphics[width=40mm]{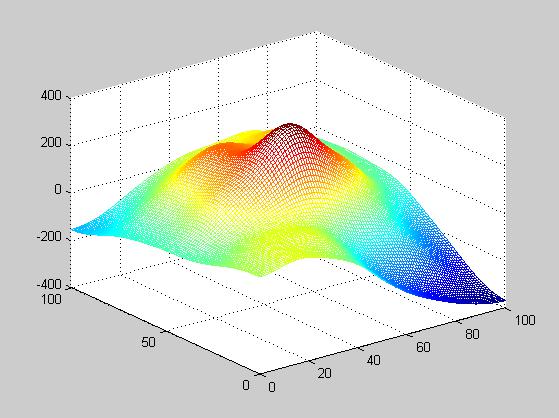}}
\subfigure[Spectral method]{
\label{FigA1d}
\includegraphics[width=40mm]{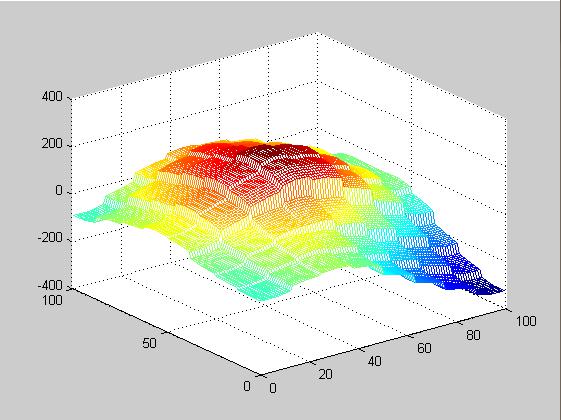}}
\centering
\caption{The potential distribution produced by a small example.}
\label{Fig:A1}
\label{fig:Aprogress}
\end{figure}

 Fig. \ref{fig:Aprogress} gives an example to demonstrate the difference between our analytical solution and the numerical solution generated by the spectral method. We take the same number of items for two series, and compare their differences. For an electrostatic system, a charge (block) is affected by all other charges (blocks). In Fig. \ref{FigA1c}, we can exactly calculate the electric potential at all coordinates in the placement region by Equations~\eqref{equ:psi}, which can reflect subtle density changes. In Fig. \ref{FigA1d}, the distribution of potential is related to the division of bin structure. The potential value of each point is not continuous and it does not reflect the details.

From the derivation process and the figure, it is obvious that our analytical solution can get a more beautiful and accurate potential distribution than the spectral method. Hence the direct solving method is more conducive to the movement of the block.

\subsection{Convergence of Analytical Solution}\label{section:convegence}

Since  $\psi(\hat{x},\hat{y})$ in Equation~(\ref{equ:psi}) is an infinite series,
we  prove that  $\psi (\hat{x},\hat{y})$ is convergent.

\begin{Lemma}
    The infinite series $\sum\limits_{p=1}^{\infty}\frac{1}{p^{3}}$ and $\sum\limits_{u=1}^{\infty}\sum\limits_{p=1}^{\infty}\frac {1}{u^3pH^{2}+up^3W^{2}}$ are convergent.
    \label{lemma1}
\end{Lemma}

\begin{Theorem}
    The infinite series $\psi(\hat{x},\hat{y})$ is absolutely convergent.
    \label{theorem1}
\end{Theorem}
\begin{proof}

Note that
 \begin{equation*}
\begin{aligned}
&-4n\leq\sum_{i=1}^{n}\left(\sin\frac {u\pi(x_i+\frac {w_i}{2})}{W}-\sin\frac {u\pi(x_i-\frac {w_i}{2})}{W}\right) \cdot\\
&~~~~~~~~~~~~~~~\left(\sin\frac {p\pi(y_i+\frac {h_i}{2})}{H}-\sin\frac {p\pi(y_i-\frac {h_i}{2})}{H}\right)\leq4n.
\end{aligned}
\end{equation*}
Then for $a_{u,p}$, $u,p\geq 1$, by Equation \eqref{equ:aup},

\begin{equation*}
|a_{u,p}|\leq\frac {4W^{2}H^{2}}{up(u^2H^{2}+p^2W^{2})\pi ^4}4n=\frac {16W^{2}H^{2}n}{up(u^2H^{2}+p^2W^{2})\pi ^4}.
  \end{equation*}
Similarly, for the other two cases, $u=0,p\geq 1$ and $u\geq 1,p=0$, we have
    $$|a_{0,p}|\le\frac{4H^2}{p^3\pi^3W}\sum_{i=1}^n w_{i}, $$
and
    $$|a_{u,0}|\le\frac{4W^2}{u^3\pi^3H}\sum_{i=1}^n h_{i}.$$
Thus
\begin{equation}
   \begin{aligned}
|\psi(\hat{x},\hat{y})|&=|\sum _{u=0}^{\infty}\sum _{p=0}^{\infty}a_{u,p}\cos\frac {u\pi\hat{x}}{W}\cos\frac {p\pi\hat{y}}{H}|\leq\sum _{u=0}^{\infty}\sum _{p=0}^{\infty}|a_{u,p}|\\
&\le\sum _{p=1}^{\infty}|a_{0,p}|+\sum _{u=1}^{\infty}|a_{u,0}|+\sum _{u=1}^{\infty}\sum _{p=1}^{\infty}|a_{u,p}|\\
&\leq\frac {4H^{2}\sum\limits_{i=1}^{n}w_{i}}{W\pi ^3}\sum\limits_{p=1}^{\infty}\frac{1}{p^{3}}+\frac {4W^{2}\sum\limits_{i=1}^{n}h_{i}}{H\pi ^3}\sum_{u=1}^{\infty}\frac{1}{u^{3}}\\
&~~~+\frac {16W^{2}H^{2}n}{\pi ^4}\sum _{u=1}^{\infty}\sum _{p=1}^{\infty}\frac {1}{u^3pH^{2}+up^3W^{2}}. \label{upper}
   \end{aligned}
  \end{equation}

By Lemma \ref{lemma1}, the above three infinite series are convergent. Hence there exists a convergent upper bound of $|\psi(\hat{x},\hat{y})|$, and thus $\psi(\hat{x},\hat{y})$ is absolutely convergent.
\end{proof}

According to Theorem \ref{theorem1}, we only need to calculate a partial sum of $\psi(\hat{x},\hat{y})$ for actual computation. Furthermore,
$\boldsymbol{\xi}(\hat{x},\hat{y})$ is the negative gradient of $\psi (\hat{x},\hat{y})$, and similarly it also can be calculated by a partial sum of $\boldsymbol{\xi}(\hat{x},\hat{y})$.
Since the denominators in Equation~\eqref{upper} contain $u^3$ or $p^3$, $\psi(\hat{x},\hat{y})$ converges quickly.
Hence we only need to take the first $K$ terms in $\psi(\hat{x},\hat{y})$ to compute a partial sum, which would be a nearly exact solution.

\subsection{Time Complexity of Calculating a Partial Sum}

In this subsection, we analyze the time complexity of calculating a partial sum of our analytical solution.
First, calculating each coefficient $a_{u,p}$ (Equations (\ref{equ:aup}), (\ref{equ:au0}), and (\ref{equ:a0p})) requires $O(n)$ time, where $n$ is the number of blocks.
Second, the electric potential $\psi (\hat{x},\hat{y})$ and the field $\boldsymbol{\xi}(\hat{x},\hat{y})$ can be computed by Equations (\ref{equ:psi}) and (\ref{equ:ksi}), respectively.
By the proof of Theorem~\ref{theorem1}, $u$ and $p$ are taken from 0 to $K$.
To obtain the partial sums of $\psi(\hat{x},\hat{y})$ and $\boldsymbol{\xi}(\hat{x},\hat{y})$, we need to calculate $K^2$ coefficients $a_{u,p}$.
Hence, computing all the coefficients $a_{u,p}$ requires $O(K^2n)$ time.

After computing all the coefficients, we must calculate $\psi (\hat{x},\hat{y})$ and $\boldsymbol{\xi}(\hat{x},\hat{y})$.
For each block $i$, $\psi_i (\hat{x},\hat{y})$ and $\boldsymbol{\xi}_i (\hat{x},\hat{y})$ denote the electric potential and the field at this block, respectively.
For $\psi_i (\hat{x},\hat{y})$ and $\boldsymbol{\xi}_i (\hat{x},\hat{y})$, we substitute $(x_i,y_i)$ into variables $(\hat{x},\hat{y})$.
According to Equations (\ref{equ:psi}) and (\ref{equ:ksi}), computing the electric potential and the field for each block requires $O(K^2)$ time.
With $n$ blocks for the problem, the total time of calculating $\psi (\hat{x},\hat{y})$ and $\boldsymbol{\xi} (\hat{x},\hat{y})$ is $O(K^2n)$.
Hence, the time complexity of computing the partial sums of the analytical solution is $O(K^2n)$.


 \subsection{Fast Computation Scheme of Poisson's Equation} \label{sec:MALM_for_GP}


Calculating a value of the partial sum needs $O(K^2n)$ time, where $K$ cannot be too small.
As a result, it is time-consuming, especially for the placement problem with millions of blocks.
In this subsection, we give a fast computation scheme of Poisson's equation based on the
analytical solution derived in Subsection~\ref{3b}.

First, we divide a chip region into $m\times m$ uniform bins,
labeled as bin $b_{lj}$, $l, j=0, 1,\cdots, m-1$. The density of bin $b_{lj}$ is
\begin{equation*}
P_{l,j}=\frac{\mbox{Total~area~of~blocks~in~bin}~b_{lj}}{\mbox{Area~of~bin}~b_{lj}}.
\end{equation*}
The area of block $i$ in bin $b_{lj}$ is determined by the block size,
which is inversely proportional to the distance between the center point of block $i$ and bin $b_{lj}$.
It is similar to the local smoothness and density scaling in ePlace.
This density expression does not differentiate between  macros and  standard cells.
Specifically, we have the following density function:
\begin{equation*}
  \rho_{l,j}(\hat{x},\hat{y})=\left\{
   \begin{aligned}
   &P_{l,j},        ~~~~~|\hat{x}-x_{l}|\le\frac{w_{b}}{2} ~~\mbox{and}~~ |\hat{y}-y_{j}|\le\frac{h_{b}}{2},  \\
   &0,~~~~~~~~~\mbox{otherwise},
   \end{aligned}
   \right.
  \end{equation*}
where $\rho_{l,j}(\hat{x},\hat{y})$ represents the global density function of bin $b_{lj}$.
Let $(x_{l},y_{j})$ denote the coordinate of the center point of bin $b_{lj}$.
Then, we accumulate the density functions of all bins, which approximates the density distribution in Equation (\ref{equ:global density}). The density function of the placement region is
 \begin{equation}
 \rho(\hat{x},\hat{y})=\sum\limits_{l=0}^{m-1}\sum\limits_{j=0}^{m-1}\rho_{l,j}(\hat{x},\hat{y})-\frac{1}{m^2}\sum\limits_{l=0}^{m-1}\sum\limits_{j=0}^{m-1}P_{l,j}.
 \label{equ:rho-dis}
 \end{equation}

Substituting Equation (\ref{equ:rho-dis}) into Equation (\ref{equation:a-b}) based on $a_{u,p}$ in Equations (\ref{equation:a-b}), (\ref{equation:a-c}), and (\ref{equation:a-d}), for the case where $u=0$ and $p\geq 1$, we get
\begin{equation}
   \begin{aligned}
&a_{0,p}\\
&=\frac{2H}{p^2\pi^2W}\iint_{R}\rho(\hat{x},\hat{y})\cos(\frac{p\pi}{H}\hat{y})d\hat{x}d\hat{y}\\
&=\frac{2H^2w_{b}}{p^3\pi^3W}\sum_{l,j=0}^{m-1}P_{l,j}\left(\sin\frac{p\pi(y_{j}+\frac{h_{b}}{2})}{H}
   -\sin\frac{p\pi(y_{j}-\frac{h_{b}}{2})}{H}\right).
   \label{equ:aupbin}
   \end{aligned}
\end{equation}
Note that $x_{l}$ and $y_j$ are determined by the bin sizes and the placement region. Substituting $x_{l}=\frac{W}{m}(l+\frac{1}{2}),y_{j}=\frac{H}{m}(j+\frac{1}{2})$ into Equation (\ref{equ:aupbin}), we get
\begin{equation*}
   \begin{aligned}
a_{0,p}&=\frac{2H^2w_{b}}{p^3\pi^3W}\sum_{l=0}^{m-1}\sum_{j=0}^{m-1}2P_{l,j}\cos(\frac{p\pi}{H}y_{j})\sin(\frac{h_b}{2}\frac{p\pi}{H})\\
&=\frac{4H^2}{p^3\pi^3m}\sin(\frac{p\pi}{2m})\sum_{l=0}^{m-1}\sum_{j=0}^{m-1}P_{l,j}\cos(\frac{p(j+\frac{1}{2})\pi}{m}).
   \end{aligned}
   \label{a0p-dis}
\end{equation*}
The other coefficients can be calculated similarly. Then, all the coefficients $a_{u,p}$ can be summarized as follows:

\begin{subnumcases}{}\label{equation:disa-a}
a_{0,0}=0,  \\\label{equation:disa-b}
a_{0,p}=\frac{4H^2}{p^3\pi^3m}\sin\frac{p\pi}{2m}\sum_{l=0}^{m-1}\sum_{j=0}^{m-1}P_{l,j}\cos\frac{p(j+\frac{1}{2})\pi}{m},\hspace{0.8cm}\\\label{equation:disa-c}
a_{u,0}=\frac{4W^2}{u^3\pi^3m}\sin\frac{u\pi}{2m}\sum_{l=0}^{m-1}\sum_{j=0}^{m-1}P_{l,j}\cos\frac{u(l+\frac{1}{2}\pi}{m}),\\
a_{u,p}=\frac{16W^2H^2}{up(u^2H^2+p^2W^2)\pi^4}\sin\frac{u\pi}{2m}\sin\frac{p\pi}{2m}\notag\\\label{equation:disa-d}
~~~~~~~~\cdot\sum_{l=0}^{m-1}\sum_{j=0}^{m-1}P_{l,j}\cos\frac{u(l+\frac{1}{2})\pi}{m}\cos\frac{p(j+\frac{1}{2})\pi}{m}.
\end{subnumcases}

For each bin $b_{lj}$, the electric potential $\psi(\hat{x},\hat{y})$ in Equation~(\ref{equ:psi}) can be recalculated by
\begin{equation*}
   \begin{aligned}
   &\psi(l,j)=\sum_{u=0}^\infty\sum_{p=0}^\infty a_{u,p}\cos(\frac{u\pi}{W}x_{l})\cos(\frac{p\pi}{H}y_{j})\\
   &~~~~~~~~=\sum_{u=0}^\infty\sum_{p=0}^\infty a_{u,p}\cos(\frac{u(l+\frac{1}{2})\pi}{m})\cos(\frac{p(j+\frac{1}{2})\pi}{m}).\\
   \end{aligned}
         \label{equ:dis-psi}
  \end{equation*}

By Theorem \ref{theorem1}, the infinite series $\psi(\hat{x},\hat{y})$ is absolutely convergent.  Thus, we can make the following approximation:
\begin{equation}
   \psi(l,j)\approx\sum_{u=0}^{K}\sum_{p=0}^{K} a_{u,p}\cos(\frac{u(l+\frac{1}{2})\pi}{m})\cos(\frac{p(j+\frac{1}{2})\pi}{m}).
            \label{equ:dis-psi1}
  \end{equation}
And the electric field $\boldsymbol{\xi}(\hat{x},\hat{y})$ in Equation (\ref{equ:ksi}) can be approximated by
\begin{equation}
  \left\{
   \begin{aligned}
   &\xi _{\hat{x}}\approx\sum_{u=0}^{K}\sum_{p=0}^{K}\frac {u\pi}{W}a_{u,p}\sin(\frac{u(l+\frac{1}{2})\pi}{m})\cos(\frac{p(j+\frac{1}{2})\pi}{m}),  \\
   &\xi _{\hat{y}}\approx\sum_{u=0}^{K}\sum_{p=0}^{K}\frac {p\pi}{H}a_{u,p}\cos(\frac{u(l+\frac{1}{2})\pi}{m})\sin(\frac{p(j+\frac{1}{2})\pi}{m}).
   \end{aligned}
   \right.
   \label{equ:dis-ksi}
\end{equation}

To make a good tradeoff between runtime and solution quality, we present the above fast computation scheme of Poisson's equation based on our analytical solution. Taking $K=m-1$ and $m=\sqrt{n}$, we can use the fast Fourier transform~\cite{RefFFTL} to quickly compute the values of Equations (\ref{equation:disa-b}), (\ref{equation:disa-c}), (\ref{equation:disa-d}), (\ref{equ:dis-psi1}), and (\ref{equ:dis-ksi}).

\begin{figure}[!htb]
\centering
\subfigure[Analytical solution]{
\label{FigB1a}
\includegraphics[width=40mm]{figure/A3.jpg}}
\subfigure[Fast computation scheme]{
\label{FigB1b}
\includegraphics[width=40mm]{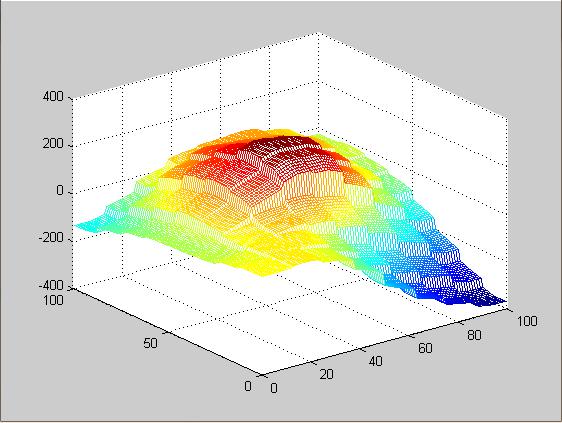}}
\subfigure[Spectral method]{
\label{FigB1c}
\includegraphics[width=40mm]{figure/A4.jpg}}
\subfigure[ Residuals of two fast methods]{
\label{FigB1d}
\includegraphics[width=40mm]{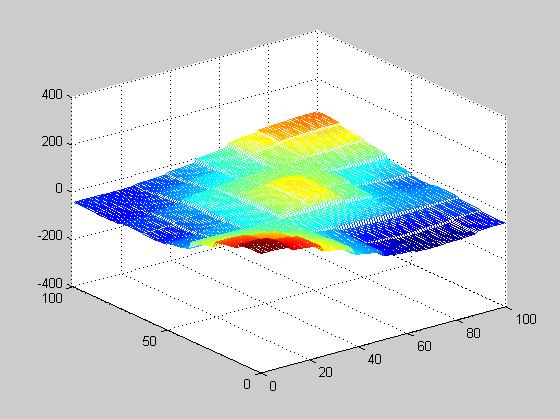}}
\centering
\caption{Differences between our method and the spectral method.}
\label{Fig:B1}
\label{fig:Bprogress}
\end{figure}

Fig. \ref{fig:Bprogress} shows the difference between our fast computation scheme and the numerical solution by the spectral method.
The density calculation and local smoothing technology in both the methods are the same. In the two images Fig. \ref{FigB1c} and Fig. \ref{FigB1d},
the overall trends are consistent. However, there are some differences between Fig. \ref{FigB1c} and Fig. \ref{FigB1d}, e.g., at the coordinate $(0,0)$. Fig. \ref{FigB1d} shows the error of  Fig. \ref{FigB1c} minus Fig. \ref{FigB1d}. According to Section \ref{3b}, our analytical solution is an accurate potential distribution, and the figure obtained by our fast computation scheme is closer to our analytical solution in detail.

We  analyze the complexity of our fast computation scheme of Poisson's equation.
In Equations (\ref{equation:disa-a})--(\ref{equation:disa-d}),
ignoring the coefficients for the summations,
we first calculate an $m\times m$ coefficient matrix of $a_{u,p}^{'}$ as follows:
\begin{subnumcases}{}
a'_{0,0}=0,  \\
a'_{0,p}=\sum_{l=0}^{m-1}\sum_{j=0}^{m-1}P_{l,j}\cos\frac{p(j+\frac{1}{2})\pi}{m},\\
a'_{u,0}=\sum_{l=0}^{m-1}\sum_{j=0}^{m-1}P_{l,j}\cos\frac{u(l+\frac{1}{2})\pi}{m},\\
a'_{u,p}=\sum_{l=0}^{m-1}\sum_{j=0}^{m-1}P_{l,j}\cos\frac{u(l+\frac{1}{2})\pi}{m}\cos\frac{p(j+\frac{1}{2})\pi}{m}.\hspace{1cm}
\end{subnumcases}
By invoking the FFT library only once, the coefficient matrix can be calculated. In addition, it takes $m^2$ time to update the coefficient matrix of $a'_{u,p}$ to $a_{u,p}$.
As a result, calculating all coefficients $a_{u,p}$ takes $O(m^2\log m^2)+O(m^2)=O(n\log n)$ time.

After calculating all the coefficients, we can calculate $\psi(l,j)$ and $\boldsymbol{\xi}(l,j)$ by using the inverse fast Fourier transform,
which
also takes $O(n\log n)$ time.
Note that $\psi(l,j)$ and $\boldsymbol{\xi}(l,j)$ are the respective electric potential and field values of bin $b_{lj}$.
According to the position of each block,
the electric potential and field values of block $i$ can be approximated by weighting the corresponding values of its surrounding bins, which takes $O(n)$ time. So the total time complexity is $O(n\log n)$.

\section{Framework of Our Pplace Algorithm}
\label{sec:our placement algorithm}

\begin{figure}[htbp] \centering
  \includegraphics[width=8.5cm]{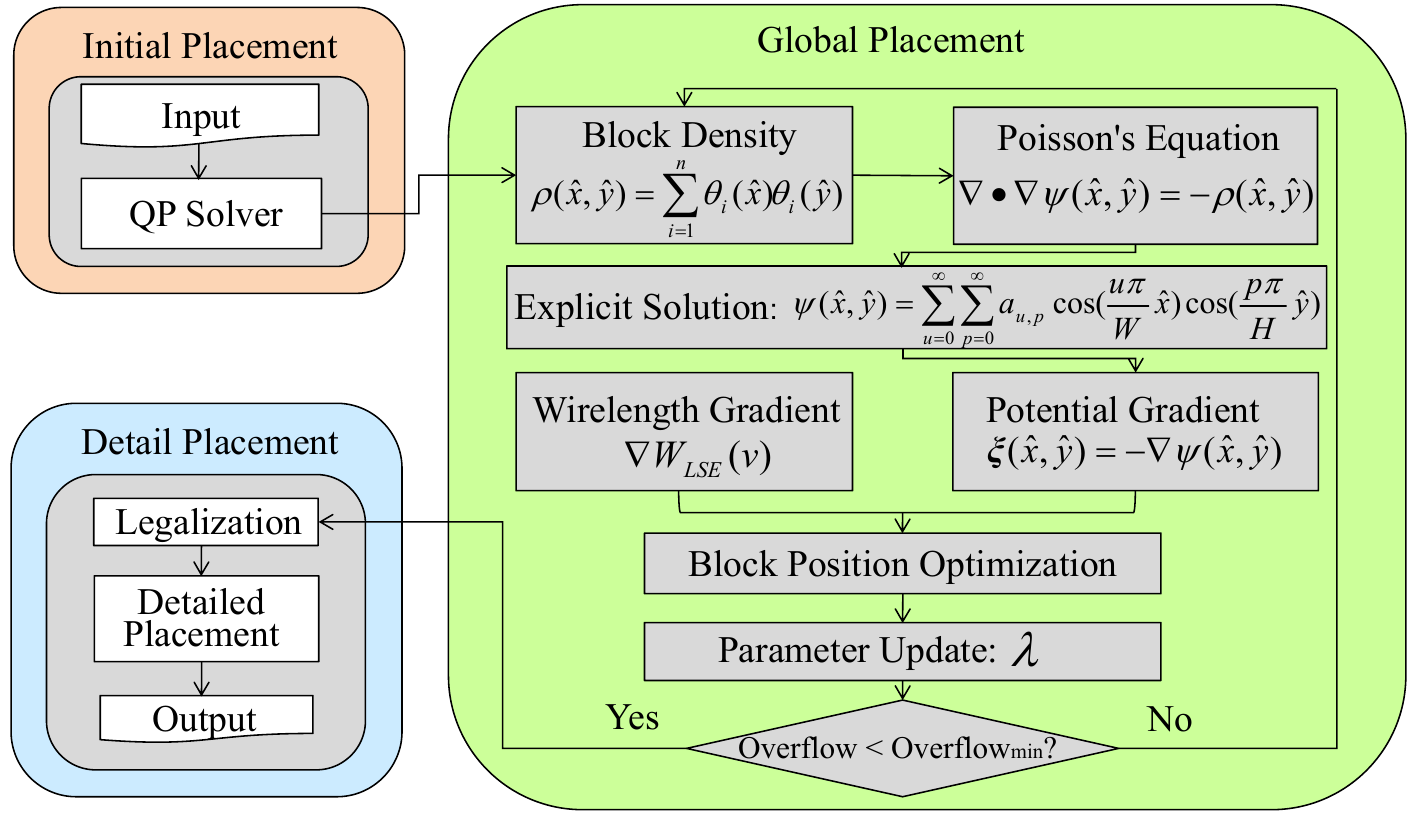}
  \caption{\small
     {Our ePlace-based Pplace placement framework}.
  }
  \label{fig:algorithm}
\end{figure}

Based on the ePlace framework, we develop our Pplace placement flow as in Figure~\ref{fig:algorithm}.
%
We reformulate the placement problem as in ePlace \cite{RefEplace}:
\begin{equation}
\operatorname*{min}\limits_{v}f(v) = W_{LSE}(v)+\lambda N(v).
\label{problem}
\end{equation}

In Problem (\ref{problem}), we consider both of the given blocks and added fillers.
Given the positions of blocks and fillers,
an exact density function can be calculated by Equation~(\ref{equ:rho}).
Using the analytical solution of Poisson's equation,
we can get the electric potential $\psi(\hat{x},\hat{y})$ and field $\boldsymbol{\xi}(\hat{x},\hat{y})$
by Equations (\ref{equ:psi}) and (\ref{equ:ksi}), respectively.

To speed up our Pplace algorithm given in Figure~\ref{fig:algorithm},
we apply the fast computation scheme of Poisson's equation
described in Subsection~\ref{sec:MALM_for_GP}.
By dividing the placement region into $m\times m$ uniform bins,
 the global density function is approximated by Equation (\ref{equ:rho-dis}).
Then, the electric potential $\psi(l,j)$ and field $\boldsymbol{\xi}(l,j)$ are approximated by Equations (\ref{equ:dis-psi1}) and (\ref{equ:dis-ksi}), respectively.
For each block, as a result,
the electric potential and field can be calculated by weighting the corresponding values of its surrounding bins.

At the $k$-th iteration, as described in Section~\ref{review ePlace},
we can quickly compute the energy (density) gradient $\nabla N^{k}=-q\boldsymbol{\xi}^k$ by using the FFT library~\cite{RefFFTL}.
With the gradient of the LSE function, we can get the gradient of the objective function $\nabla f(v^{k})$.

For fair comparison between the numerical solution method in ePlace and our fast computation scheme of Poisson's equation,
we use the same optimization method and parameters as in ePlace.
At each iteration, as in ePlace, we use Nesterov's method to solve the unconstrained minimization problem
by only one iteration and get a new solution $(x^{k+1}, y^{k+1})$,
and then update the penalty parameter $\lambda$.
If the overflow ratio meets the termination criterion,
then the algorithm is terminated.
After obtaining a global placement result,
the legalization and detailed placement solvers provided by NTUplace3 \cite{RefNtu3}
are called to obtain a final placement result.

\section{Experimental Results} \label{sec:experimental results}

In this section, we first compare our proposed Pplace placer with ePlace.
Then, we embed the density calculation method into NTUplace3 by replacing its density control method to show our robustness.
We implemented our Pplace placer using the C++ programming language and ran the program in a single-thread mode on a Linux machine with 3.20GHz Intel Core i5 6500 CPU and 16GB memory.
In our experiments, we used the ISPD 2005~\cite{Ref2005} and ISPD 2006 benchmark suites~\cite{Ref2006} for our comparative studies based on
wirelength and runtime \emph{because it is easier to see the effects of the algorithms with these fundamental placement metrics}.
Note that we do not use more recent placement contest benchmarks because these contests were mainly on routability optimization with various constraints (e.g., fence regions); further
ePlace was reported to achieve the best published wirelength based on these benchmarks.

Table \ref{table:exp_result} and Table \ref{table:exp_result.06} show the circuit statistics of the ISPD 2005 and ISPD 2006 benchmarks,
for which the problem sizes range from 211K to 2508K.
The numbers of blocks, nets, standard cells, macros and target density for the benchmark are denoted by ``\#Blocks'', ``\#Nets'', ``\#Std. Cells'', ``\#Mac.'' and ``\#Density'', respectively.

\subsection{Comparison with ePlace}

In the ePlace work~\cite{RefEplace}, it shows that ePlace achieves the shortest total wirelength for all the eight benchmarks compared with ten academic placers. In addition, it is the fastest among all nonlinear (non-quadratic) placers.
For fair comparison on \emph{global placement},
we use the legalization and detailed placement methods by NTUplace3~\cite{RefNtu3} in our placer,
the same as in ePlace.

Since we only have the ePlace binary code, but do not have its source code,
it is not feasible to directly compare the numerical method in ePlace with our fast computation scheme of Poisson's equation
based on the ePlace version in~\cite{RefEplace}.
As a result, we implemented the numerical solution by spectral method and named it the ``num-spectral''.

Using the same legalization and detailed placement methods,
the HPWL and runtime comparisons among ePlace, ``num-spectral'' and our Pplace placer are reported in Table~\ref{table:exp_result} and Table~\ref{table:exp_result.06}, respectively.
In the tables, ``GP-HWPL'' gives the HPWL wirelength after \emph{global placement},
``GP-CPU'' the corresponding runtime,
``HWPL'' the HPWL wirelength of the final placement, ``sHWPL'' the scaled HPWL,
and ``CPU (s)'' the runtime in seconds of the total running time of a placer.
The bottom row gives the normalized wirelength and runtime ratios based on our results,
and the best results among the three placers are marked in bold.

\begin{table*}[htbp]
  \centering
  \caption{Comparisons among ePlace, ``num-spectral,'' and ours based on the ISPD 2005 Contest benchmarks~\cite{Ref2005}.}
  \scalebox{0.6}[1]{
  \vspace{0.0cm}
  \begin{tabular}{|c|c|c|c|c|c|c|c|c|c|c|c|c|c|c|c|c| }
  \hline
       \multicolumn{1}{|c|}{\empty} &
       \multicolumn{4}{c|}{Benchmark Statistics} &
       \multicolumn{4}{c|}{ePlace} &
       \multicolumn{4}{c|}{``num-spectral''} &
       \multicolumn{4}{c|}{Our Pplace}
       \\ \cline{2-17}
  Benchmark &\#Blocks &\#Nets &\#Std. Cells &\#Mac. &GP-HPWL&GP-CPU&HPWL &CPU (s)&GP-HPWL&GP-CPU&HPWL &CPU (s)&GP-HPWL&GP-CPU&HPWL &CPU (s)\\
  \hline
   adaptec1    &211447  &221142  &210904  &543  &73118044&135 &74643953          &164           &73604755&110&75989399  &\bfseries132  &72148076&120&\bfseries73387186   &147   \\
   adaptec2    &255023  &266009  &254457  &566  &83534323&187 &84863690          &228           &81206401&156&85600480  &200  &80249273&158&\bfseries83033741   &\bfseries195   \\
   adaptec3    &451650  &466758  &450927  &723  &193730691&612 &196395438         &688           &191055370&448&198117119 &\bfseries531  &189583003&567&\bfseries193577147  &640  \\
   adaptec4    &496045  &515951  &494716  &1329 &178102879&657 &179148257         &738           &181105154&569&184729214 &627  &174548043&448&\bfseries177299566  &\bfseries524   \\
   bigblue1    &278164  &284479  &277604  &560  &89625096&262 &90985255          &\bfseries300  &92122364&275&91608281  &315  &89133405&298&\bfseries89739424   &332            \\
   bigblue2    &557866  &577235  &534782  &23084&139657694&371 &141819172         &\bfseries495  &141215231&433&146551686 &537  &137121253&438&\bfseries140828061  &548            \\
   bigblue3    &1096812 &1123170 &1093034 &3778 &296504804&1143 &308004959         &1330          &311820347&870&311618882 &\bfseries1061 &300830027&1013&\bfseries306369240  &1204  \\
   bigblue4    &2177353 &2229886 &2169183 &8170 &739782972&2300 &756310869&\bfseries2804 &772231562&3386&774801191 &3891 &744151236&2817&\bfseries754594872           &3187           \\ \hline
   Normalized  &        &        &        &     &1.01    &1.07    &1.01             &1.07         &1.03&0.99&1.03     &0.99&1.00&1.00&1.00                  &1.000      \\

  \hline
  \end{tabular}
  \label{table:exp_result}
  }
\end{table*}

Compared with ePlace, from  Table~\ref{table:exp_result}, our Pplace placer achieves better HPWL results on all the eight benchmarks.
On average, Pplace achieves 1\% smaller final HPWL and is 7\% faster than ePlace.
Pplace also achieves 3\% smaller HPWL than ``num-spectral'' in almost the same running time.
Our HPWL improvement over ``num-spectral'' justifies the effectiveness of our fast computation scheme for the
analytical solution of Poisson's equation, because the only difference between ``num-spectral'' and Pplace
lies in the solution methods of Poisson's equation
(i.e., a numerical method by the spectral method and an analytical solution for ours).
In particular, the improved normalized values for global placement ``GP-HWPL'' and for final placement ``HPWL''
are almost the same, revealing that our algorithm is effective and efficient for global placement optimization.

\begin{figure}[!htb]
\centering
\subfigure[Iterations=0]{
\label{Fig1a}
\includegraphics[width=30mm]{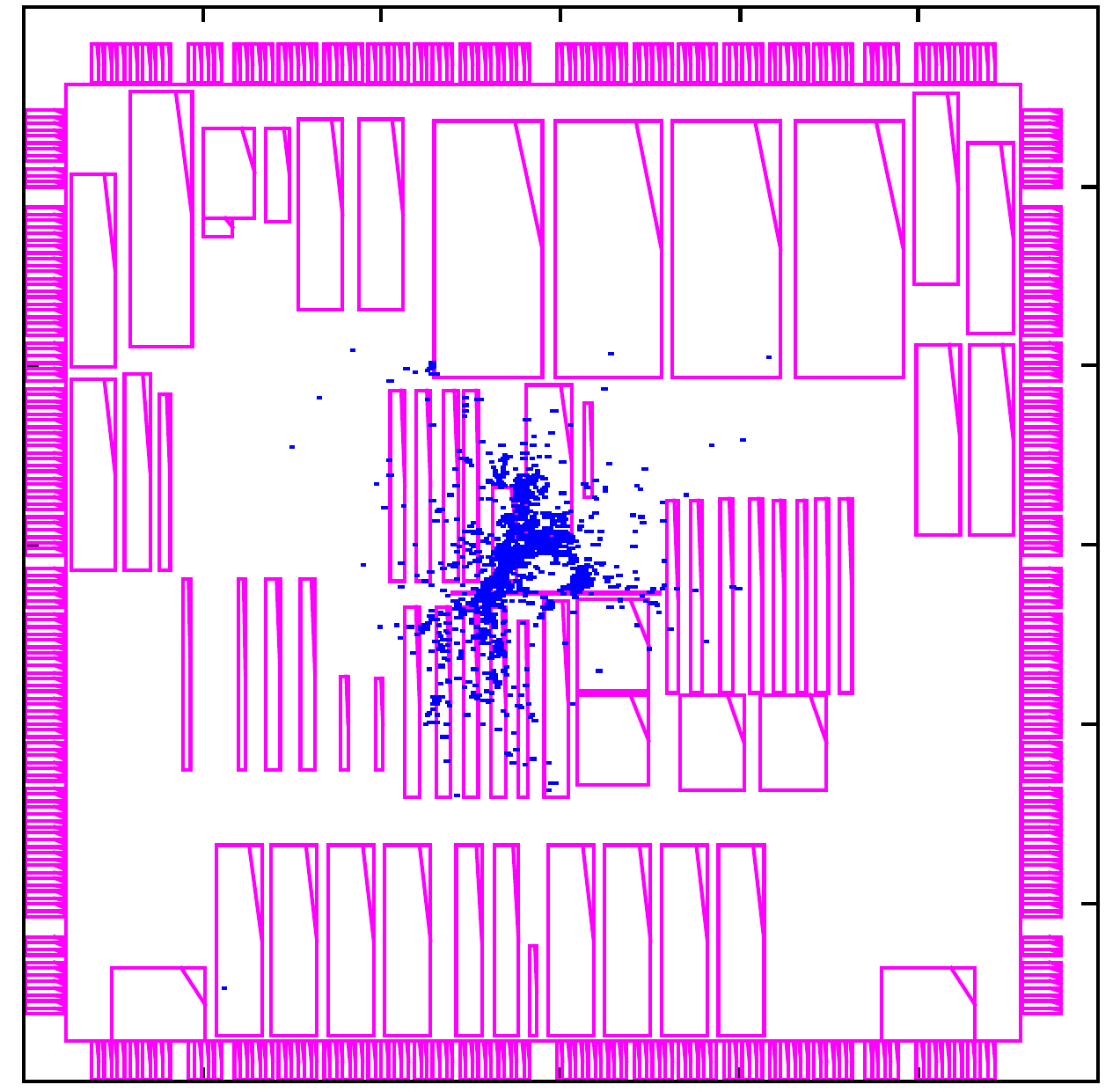}}
\hspace{0.5cm}
\subfigure[Iterations=120]{
\label{Fig1b}
\includegraphics[width=30mm]{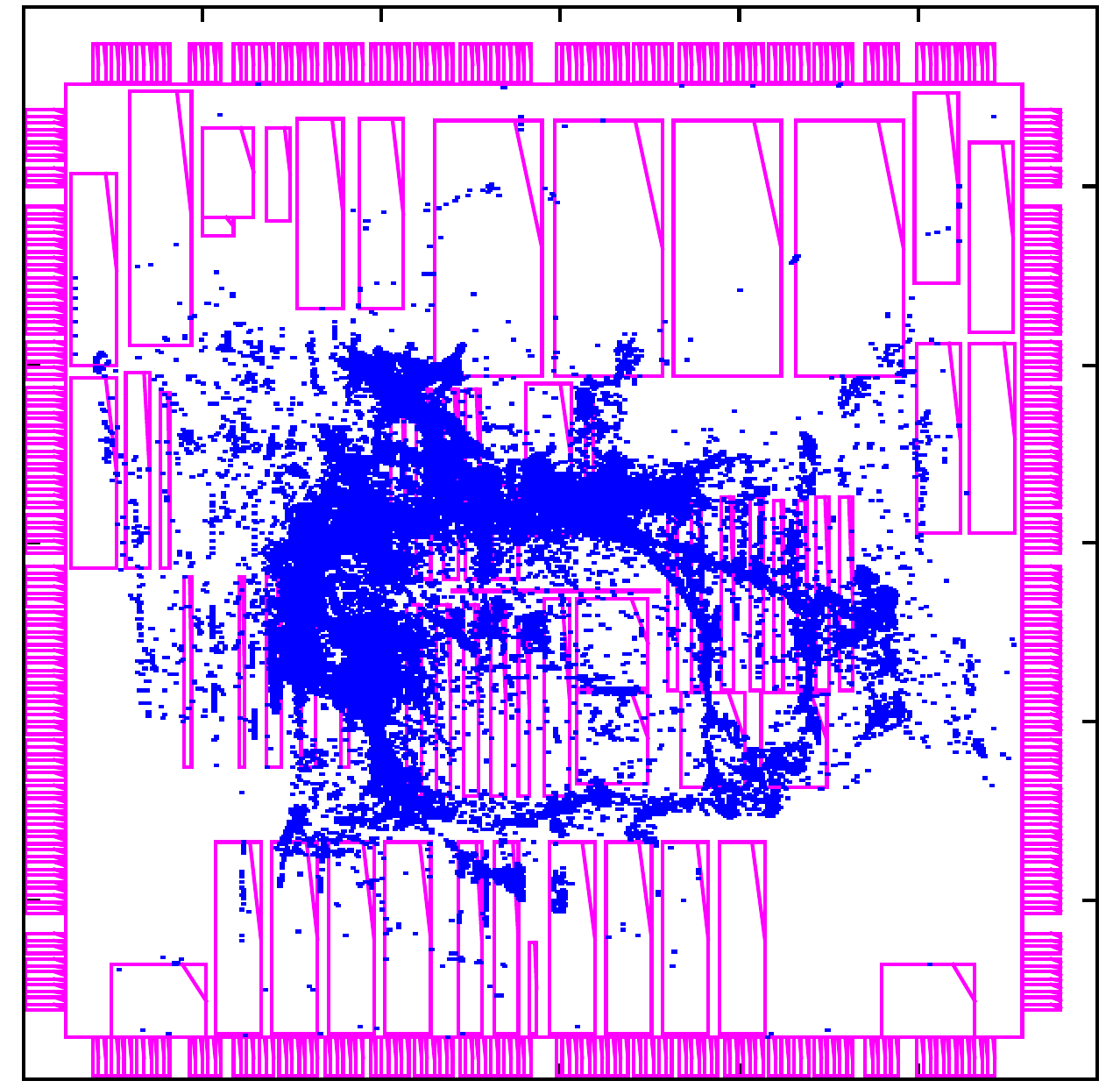}}
\hspace{0.5cm}
\subfigure[Iterations=240]{
\label{Fig1c}
\includegraphics[width=30mm]{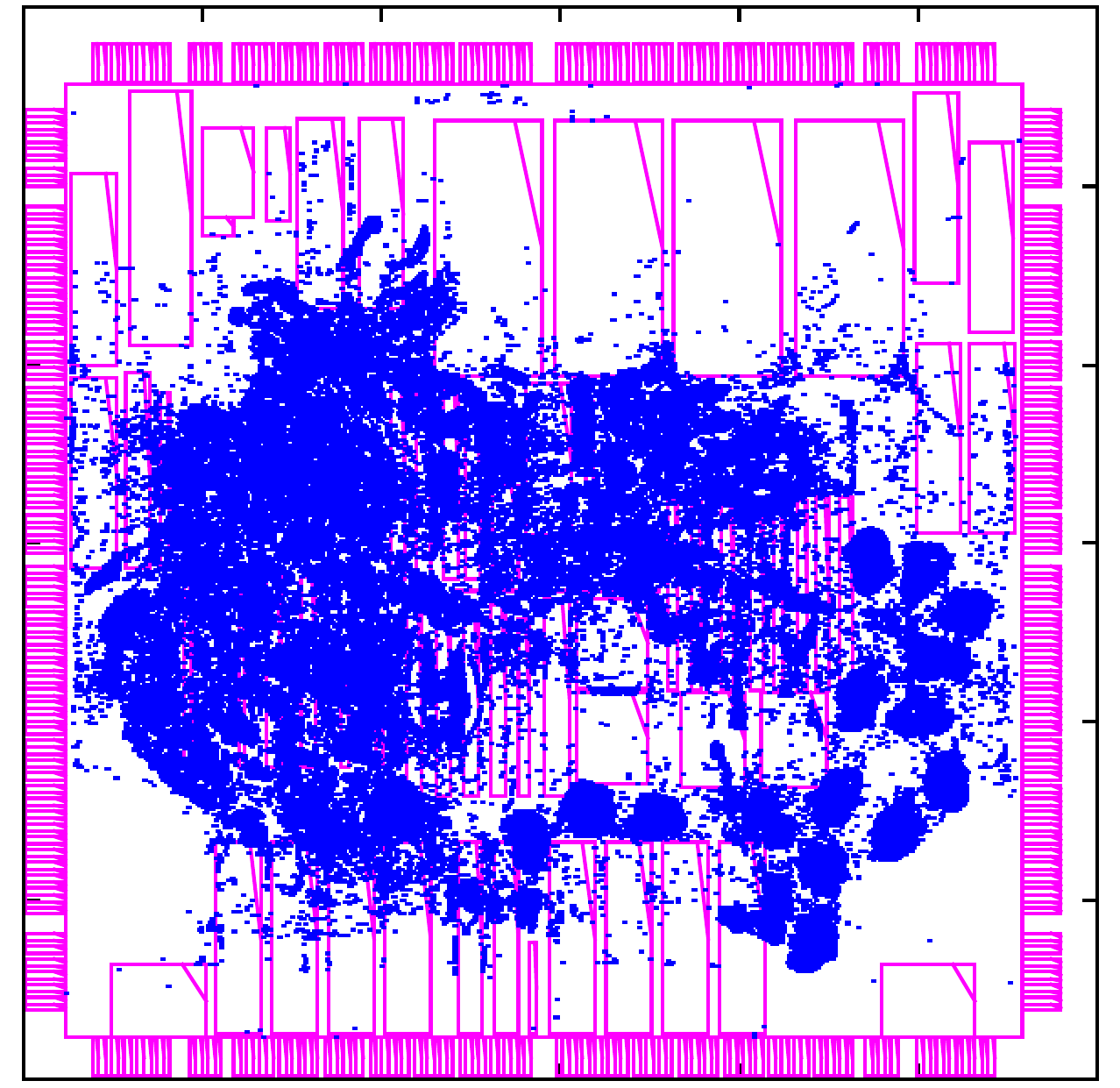}}
\hspace{0.5cm}
\subfigure[Iterations=358]{
\label{Fig1d}
\includegraphics[width=30mm]{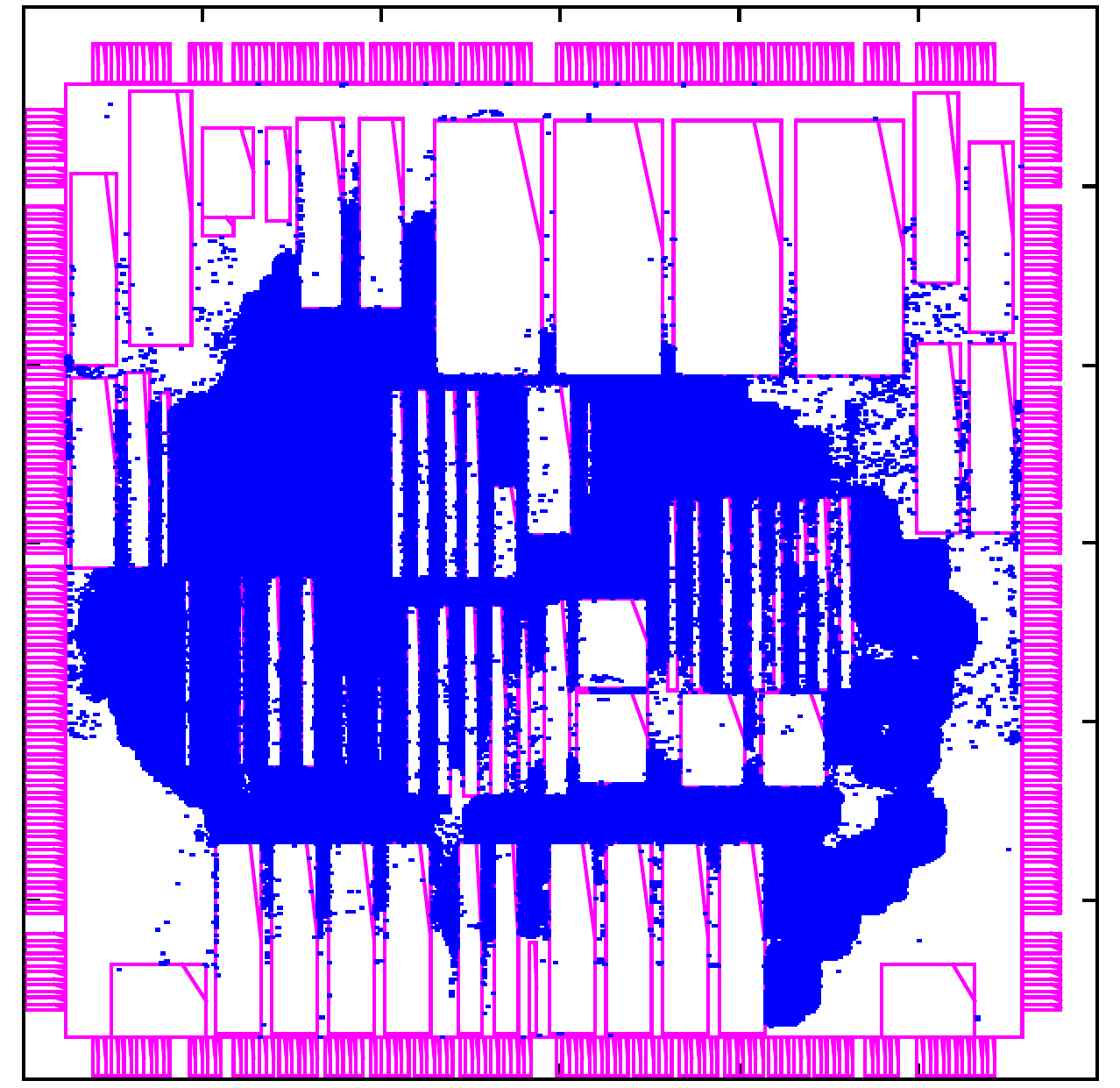}}
\centering
\caption{Blocks distribution during our global placement progression for the adaptec1 benchmark.
Standard cells and macros are denoted by blue points and pink rectangles, respectively.}
\label{Fig:1}
\label{fig:progress}
\end{figure}

Figure~\ref{fig:progress} illustrates the block distribution during global placement by Pplace on the adaptec1 benchmark.
Note that fillers are not shown in order to better observe the movements of blocks (marked in blue).

We also tested our algorithm on the ISPD 2006 benchmarks \cite{Ref2006}, and compare the results with ePlace in Table \ref{table:exp_result.06}.
The experimental results show that, our Pplace placer also achieves better HPWL and sHPWL  on all the eight benchmarks.
On average, Pplace achieves respectively 1\% and 1\% shorter HPWL and sHPWL than ePlace.
\begin{table*}[htbp]
  \centering
  \caption{Comparisons among ePlace, ``num-spectral,'' and ours based on the ISPD 2006 Contest benchmarks~\cite{Ref2006}.}
  \scalebox{0.7}[1]{
  \vspace{0.0cm}
  \begin{tabular}{|c|c|c|c|c|c|c|c|c|c|c|c|c|c| }
  \hline
       \multicolumn{1}{|c|}{\empty} &
       \multicolumn{4}{c|}{Benchmark Statistics} &
       \multicolumn{3}{c|}{ePlace} &
       \multicolumn{3}{c|}{``num-spectral''} &
       \multicolumn{3}{c|}{Our Pplace}
       \\ \cline{2-14}
  Benchmark &\#Blocks &\#Std. Cells &\#Mac. &\#Density &HPWL&sHPWL &CPU (s)&HPWL&sHPWL &CPU (s)&HPWL&sHPWL &CPU (s)\\
  \hline
   adaptec5    &843128  &842482  &646  &50\%   &394695282 &397557738&\bfseries1265 &403286002&406217085 &1266 &392207460 &\bfseries397487761 &1466   \\
   newblue1    &330474  &330037  &401  &80\%   &62304524  &62451418 &367  &62448920 &62922805  &\bfseries245  &58129002  &\bfseries58585324  &323   \\
   newblue2    &441516  &436516  &5000  &90\%  &181378387 &182354956&376  &186214953&188717090 &\bfseries353  &181379584 &\bfseries182182053 &396  \\
   newblue3    &494011  &482833  &11178  &80\% &265770119 &265991914&588  &266567897&269089963 &609  &263713231 &\bfseries264092804 &\bfseries565   \\
   newblue4    &646139  &642717  &3422  &50\%  &272832375 &276140253&770  &261597190&286468349 &786  &264451809 &\bfseries268925895 &\bfseries679            \\
   newblue5    &1233058  &1228177  &4881  &50\%&489738331 &492808559&1978 &485240543&497299421 &2200 &484023312 &\bfseries489359921 &\bfseries1801            \\
   newblue6    &1255039 &1248150 &6889 &80\%   &462673606 &464468022&\bfseries1919 &472367792&474858888 &2441 &462220925 &\bfseries463979824 &2416  \\
   newblue7    &2507954 &2481372 &26582 &80\%  &987346891 &989865614&\bfseries3246 &986219208&1007701401&4705 &987060153 &\bfseries989058803&3670           \\ \hline
   Normalized  &        &        &        &    &1.01     &1.01    &0.99&1.02    &1.03     &1.05&1.00&1.00                 &1.00      \\

  \hline
  \end{tabular}
  \label{table:exp_result.06}
  }
\end{table*}


\subsection{Robustness of Our Fast Computation Scheme}

NTUPlace3~\cite{RefNtu3} is a high-quality, robust placer with a multilevel framework.
To examine the robustness of our method,
we replace the density control in NTUplace3 with our fast computation scheme of Poisson's equation.
Also, the density gradient calculation in NTUPlace3 is changed to the electric field.
For convenience, the modified NTUplace3 is called MNTU3-multilevel.
Table~\ref{table:exp_result1} reports the experimental results of NTUPlace3 and MNTU3-multilevel.

\begin{table}[htbp]
  \centering
  \caption{Comparisons of the HPWL and Runtime (seconds) among NTUplace3, MNTU3-multilevel, and MNTU3-flat.}
  \scalebox{0.6}[1]{
  \vspace{0.0cm}
  \begin{tabular}{|c|c|c|c|c|c|c|c|c| }
  \hline
       \multicolumn{1}{|c|}{\empty} &
       \multicolumn{2}{c|}{NTUplace3} &
       \multicolumn{2}{c|}{MNTU3-multilevel}&
       \multicolumn{2}{c|}{MNTU3-flat }
       \\ \cline{2-7}
  Benchmark         &HPWL  &CPU(s) &HPWL &CPU(s)&HPWL  &CPU(s) \\
  \hline
   adaptec1    &80291340  &183           &74860458  &\bfseries125    &\bfseries74696137   &137 \\
   adaptec2    &90180949  &226           &85048037  &\bfseries163    &\bfseries83314442   &173 \\
   adaptec3    &233773979 &\bfseries501  &201084986 &530    &\bfseries196870086  &592\\
   adaptec4    &215015818 &602           &181371097 &554    &\bfseries178493350  &\bfseries520\\
   bigblue1    &98653718  &349           &91868762  &\bfseries239    &\bfseries89746505   &322 \\
   bigblue2    &158269695 &876           &144905354 &461    &\bfseries144599935  &\bfseries399\\
   bigblue3    &346325425 &1044 &327982683 &\bfseries862  &\bfseries311977354  &1123\\
   bigblue4    &829086990 &3125 &817964751 &\bfseries2235  &\bfseries799215445           &2347 \\  \hline
   Normalized  &1.11   &1.27          &1.02   &0.93  &1.00             &1.00 \\
  \hline
  \end{tabular}
  \label{table:exp_result1}
  }
\end{table}

From the results, MNTU3-multilevel achieves 9\% smaller HPWL and 37\% shorter time than NTUplace3.
These results not only justify the effectiveness of our fast computation scheme of Poisson's equation,
but also the robustness of our analytical solution of Poisson's equation for a different placer.

To further demonstrate the robustness of our solution of Poisson's equation,
we turned off the multilevel clustering in MNTU3-multilevel and named this method MNTU3-flat.
From Table~\ref{table:exp_result1}, MNTU3-flat achieves 11\% smaller HPWL and 27\% shorter runtime than NTUplace3,
and 2\% smaller HPWL than MNTU3-multilevel in 7\% longer runtime.
The results again shows the robustness of our solution of Poisson's equation.



\section{Conclusions} \label{sec:conclusion}

Unlike previous global placement methods that solve Poisson's equation numerically, we have derived an analytical solution and a fast computation scheme of Poisson's equation for distributing circuit blocks effectively and efficiently.
The analytical solution is an infinite series, which converges absolutely.
Nevertheless, calculating a partial sum of the analytical function takes $O(K^2n)$ time.
In order to quickly solve Poisson's equation,
we have divided the placement region into uniform bin grids,
and used the fast Fourier transform to approximately computing the analytical solution.
Using the fast computation scheme of Poisson's equation,
we have developed the new placer Pplace.
Experimental results on the ISPD 2005 and ISPD 2006 benchmarks have shown that Pplace with the fast computation scheme of Poisson's equation is effective, efficient, and robust.
Future work should be extending the methods in this paper to consider the VLSI placement problem with different constraints, e.g., routability. Moreover, it would be interesting using the solution method in this paper to other partial differential equations for EDA and other applications.
With the pervasive applications of Poisson's equation in scientific fields, in particular,
our effective, efficient, and robust computation scheme for its analytical solutions
can provide substantial impacts on these fields.


\bibliographystyle{abbrv}
\bibliography{sigproc}


\end{document}